\documentclass[twocolumn,showpacs,superscriptaddress,prb,amsmath,amssymb]{revtex4}

\usepackage{graphicx}
\usepackage{dcolumn}
\usepackage{bm}
\usepackage{epstopdf}
\usepackage{epsfig}

\begin{document}

\title{Fragile singlet ground state magnetism in pyrochlore osmates $R_2$Os$_2$O$_7$ ($R$=Y and Ho)}

\author{Z. Y. Zhao}
\affiliation{Department of Physics and Astronomy, University of Tennessee, Knoxville, Tennessee 37996, USA}
\affiliation{Materials Science and Technology Division, Oak Ridge National Laboratory, Oak Ridge, Tennessee 37831, USA}

\author{S. Calder}
\affiliation{Quantum Condensed Matter Division, Oak Ridge National Laboratory, Oak Ridge, Tennessee 37831, USA}

\author{A. A. Aczel}
\affiliation{Quantum Condensed Matter Division, Oak Ridge National Laboratory, Oak Ridge, Tennessee 37831, USA}

\author{M. A. McGuire}
\affiliation{Materials Science and Technology Division, Oak Ridge National Laboratory, Oak Ridge, Tennessee 37831, USA}

\author{B. C. Sales}
\affiliation{Materials Science and Technology Division, Oak Ridge National Laboratory, Oak Ridge, Tennessee 37831, USA}

\author{D. G. Mandrus}
\affiliation{Materials Science and Technology Division, Oak Ridge National Laboratory, Oak Ridge, Tennessee 37831, USA}
\affiliation{Department of Materials Science and Engineering, University of Tennessee, Knoxville, Tennessee 37996, USA}

\author{G. Chen}
\affiliation{Collaborative Innovation Center of Advanced Microstructures and Department of Physics and Center for Field Theory and Particle Physics, Fudan University, Shanghai 200433, China}

\author{N. Trivedi}
\affiliation{Department of Physics, The Ohio State University, Columbus, Ohio 43210, USA}

\author{H. D. Zhou}
\affiliation{Department of Physics and Astronomy, University of Tennessee, Knoxville, Tennessee 37996, USA}

\author{J.-Q. Yan}
\affiliation{Materials Science and Technology Division, Oak Ridge National Laboratory, Oak Ridge, Tennessee 37831, USA}
\affiliation{Department of Materials Science and Engineering, University of Tennessee, Knoxville, Tennessee 37996, USA}

\date{\today}

\begin{abstract}

The singlet ground state magnetism in pyrochlore osmates Y$_2$Os$_2$O$_7$ and Ho$_2$Os$_2$O$_7$ is studied by DC and AC susceptibility, specific heat, and neutron powder diffraction measurements. Despite the expected non-magnetic singlet in the strong spin-orbit coupling (SOC) limit for Os$^{4+}$ ($5d^4$), Y$_2$Os$_2$O$_7$ exhibits a spin-glass (SG) ground state below 4 K with weak magnetism, suggesting possible proximity to a quantum phase transition between the non-magnetic state in the strong SOC limit and the magnetic state in the strong superexchange limit. Ho$_2$Os$_2$O$_7$ has the same structural distortion as occurs in Y$_2$Os$_2$O$_7$. However, the Os sublattice in Ho$_2$Os$_2$O$_7$ shows long-range magnetic ordering below 36\,K. The sharp difference of the magnetic ground state between Y$_2$Os$_2$O$_7$ and Ho$_2$Os$_2$O$_7$ signals the singlet ground state magnetism in $R_2$Os$_2$O$_7$ is fragile and can be disturbed by the weak $4f-5d$ interactions.

\end{abstract}


\maketitle

\section{Introduction}

Singlet ground state magnetism, sometimes also named induced magnetism, was initially studied in rare earth compounds in 1960s. \cite{Cooper} In, for example,  Pr$_3$In, Pr$_3$Tl, and Tb$P$ ($P$ = P, As, Sb, Bi), where the crystal-field-only ground state of the rare earth ions is a singlet, magnetic ordering at zero temperature occurs as the ratio of exchange to crystal-field interaction is above a certain critical value. This magnetic ordering out of singlets takes place through a polarization instability of the crystal-field-only singlet ground state, which distinguishes itself from the normal cases where permanent moments are aligned below the magnetic ordering temperature. In transition metal compounds, transition metal ions can sometimes be stabilized in a singlet state in the presence of metal-metal dimerization, particular crystal field splitting, strong positive single-site anisotropy,  and strong spin-orbit coupling (SOC). \cite{Daniel}

In the presence of strong SOC, the three $t_{2g}$ orbitals entangle with the spins and form an upper $J_{\text{eff}}=1/2$ doublet and a lower $J_{\text{eff}} =3/2$ quadruplet. This J$_{eff}$ picture has been employed to understand the SOC assisted Mott insulating state of Sr$_2$IrO$_4$.\cite{Kim_Sr2IrO4} Following this $J_{\text{eff}}$ scenario, the local moment of a $d^4$ electron configuration in an octahedral crystal field is a trivial singlet with four electrons filling the lower quadruplet. This state is also named $J_{\text{eff}} =0$ nonmagnetic state. Indeed, many systems with $5d^4$ configuration, such as NaIrO$_3$ \cite{NaIrO3} and \textit{A}$_2$\textit{R}IrO$_6$ ($A$ = Sr, Ba, $R$ = Sc, La, Lu, Ho) \cite{A2RIrO6,Ba2RIrO6} with Ir$^{5+}$ ions, and $A$OsO$_3$ ($A$ = Ca, Sr, Ba) with Os$^{4+}$ ions, \cite{BaOsO3} exhibit non-magnetic ground states.

However, novel magnetism governed by gapped singlet-triplet excitations was proposed theoretically. \cite{Khaliullin} A modest Hund's coupling, J$_H$, would change the antiferromagnetic superexchange at J$_H$=\,0 to a ferromagnetic interaction; superexchange interactions competes with SOC and induces magnetic transitions between the J=0 nonmagnetic state and J=1 or 2 ferromagnetic states.\cite{Nandini} The magnetic ordering observed in a double perovskite Sr$_2$YIrO$_6$ with Ir$^{5+}$ ($5d^4$) ions is the first example of the novel magnetism out of a singlet ground state predicted by atomic physics in the presence of strong SOC. \cite{Cao_SrYIrO} In Sr$_2$YIrO$_6$, the IrO$_6$ octahedra are flattened with a shorter bond length between Ir and the apical oxygen. The observed magnetism was argued to result from the noncubic crystal field and its interplay with local exchange interactions and SOC.

Most recent studies observed magnetic moments in cubic Ba$_2$YIrO$_6$ with a space group Fm-3m which does not allow trigonal distortion, tilting, or rotation  of IrO$_6$ octahedra.\cite{Ba2YIrO6} And  this magnetic moment shows little dependence on the Ba/Sr ratio in Ba$_{2-x}$Sr$_x$YIrO$_6$ where the IrO$_6$ octahedral distortion increases with Sr substitution. \cite{Cava} All these observations signal the importance of SOC in determining the magnetism of d$^4$ compounds and support the theoretically predicted excitonic magnetism\cite{Khaliullin, Nandini} if J is a valid quantum number. \cite{Breakdown_J0}

In this work, we report the singlet ground state magnetism in pyrochlore osmates Y$_2$Os$_2$O$_7$ and Ho$_2$Os$_2$O$_7$ in which the Os$^{4+}$ ion has a $5d^4$ electron configuration. The pyrochlore osmates are scarcely studied with only one report in 1970s. \cite{R2Os2O7} Our neutron diffraction study reveals that the OsO$_6$ octahedra show significant trigonal distortion in both Y$_2$Os$_2$O$_7$ and Ho$_2$Os$_2$O$_7$. Thus $R_2$Os$_2$O$_7$ is another model system for the study of singlet ground state magnetism and unusual quantum phase transition \cite{Nandini} induced by intersite superexchange and trigonal crystal fields  in the presence of strong SOC. Despite the same structural distortion, Y$_2$Os$_2$O$_7$ develops a spin-glass (SG) state with weak magnetism below 4 K, while Ho$_2$Os$_2$O$_7$ shows a long-range magnetic ordering below 36 K. The sharp difference between the magnetic ground states of Y$_2$Os$_2$O$_7$ and Ho$_2$Os$_2$O$_7$ signals that the singlet ground state magnetism in \emph{R}$_2$Os$_2$O$_7$ is fragile and can be disturbed by other factors, such as the weak $f-d$ interaction between Ho$^{3+}$ and Os$^{4+}$ moments in Ho$_2$Os$_2$O$_7$.

\begin{figure}
\includegraphics[clip,width=8.5cm]{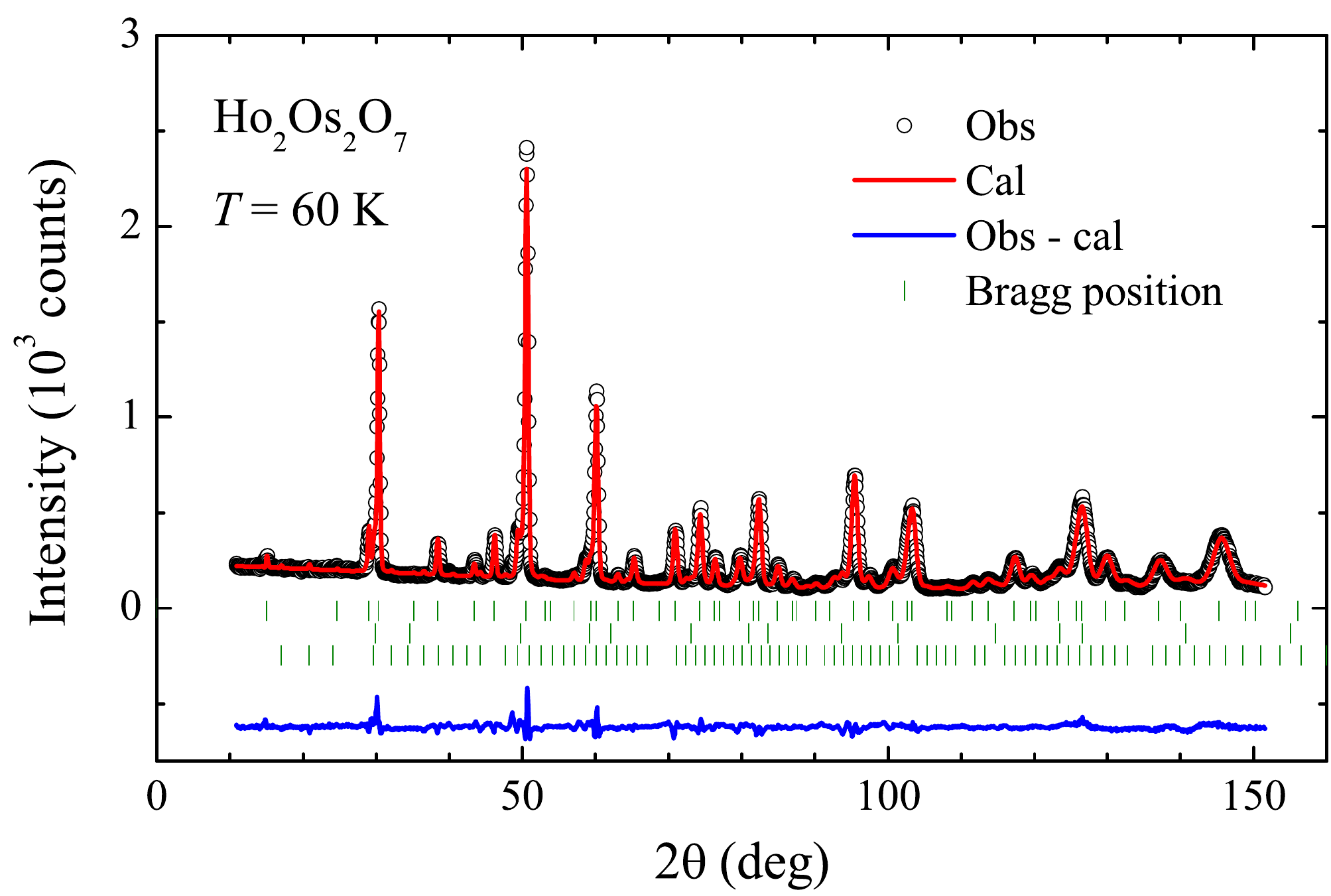}
\caption{(Color online) Neutron powder diffraction pattern of Ho$_2$Os$_2$O$_7$ collected at 60 K on HB-2A with $\lambda$ = 1.54 {\AA}. The red line on the top is the refined pattern including three phases. The vertical ticks are the Bragg positions of Ho$_2$Os$_2$O$_7$ ($Fd\bar3m$), Ho$_2$O$_3$ ($Fm\bar3m$), and Ho$_2$O$_3$ ($Ia\bar3$) phases, respectively from top to bottom. The blue line on the bottom is the difference between the observed and calculated intensities.}
\label{NPD}
\end{figure}

\section{Experiments}

Polycrystalline Y$_2$Os$_2$O$_7$ and Ho$_2$Os$_2$O$_7$ were synthesized by the solid state reaction method. Y$_2$O$_3$ and Ho$_2$O$_3$ were dried at 950$^{\circ}$C overnight before using. A homogeneous mixture of Y$_2$O$_3$ (or Ho$_2$O$_3$) and Os was loaded into an alumina crucible with an appropriate mount of AgO placed in another crucible as the source of oxygen. Both crucibles were sealed in a quartz tube and reacted at 1000$^{\circ}$C for 4 days. The powder was then reground, pelletized, and fired in a sealed quartz tube at 1000$^{\circ}$C for another 4 days. The DC magnetic properties were measured between 2-300 K using a Quantum Design (QD) magnetic property measurement system. Specific heat, electrical resistivity, and AC susceptibility were measured between 2-200 K and 2-60 K using a QD physical property measurement system, respectively. Neutron powder diffraction was performed at High Flux Isotope Reactor (HFIR), Oak Ridge National Laboratory (ORNL). The local structural distortion was studied using HB-2A powder diffractometer with $\lambda$ = 1.54 {\AA} for both compounds. Diffraction study with $\lambda$ = 2.41 {\AA} was  performed to look for possible magnetic reflections. Y$_2$Os$_2$O$_7$ was also measured using HB-1A fixed-incident-energy triple-axis spectrometer with $\lambda$ = 2.36 {\AA} which has a higher flux and cleaner background.

\section{Results and Discussion}

The synthesis of polycrystalline \textit{R}$_2$Os$_2$O$_7$ (\textit{R}=Y and Ho) by solid state reaction has been found to be rather challenging. A complete reaction between the starting materials has never been reached. The best sample we have made contains 3-5\% starting materials as impurities. Figure \ref{NPD} shows the neutron powder diffraction pattern measured at 60 K for Ho$_2$Os$_2$O$_7$ used in this study. The extra reflections come from Ho$_2$O$_3$ including about 10\% $Fm\bar3m$ and 5\% $Ia\bar3$ phases. The Y$_2$Os$_2$O$_7$ powder used in this study has 7\% Y$_2$O$_3$ and 10\% OsO$_2$ determined from neutron powder diffraction. The paramagnetic Ho$_2$O$_3$ and OsO$_2$ and non-magnetic Y$_2$O$_3$ are not expected to affect the determination of the magnetic ordering temperatures. The refinement of both samples suggests the majority phase is stoichiometric, and the lattice parameters are consistent with the previous report. \cite{R2Os2O7} The refined structural parameters at different temperatures for both compositions are summarized in Table 1.

\begin{table*}[!ht]
\caption{Structural parameters of Y$_2$Os$_2$O$_7$ and Ho$_2$Os$_2$O$_7$ with space group $Fd\bar{3}m$ measured using $\lambda$ = 1.54 {\AA}}
\label{table1}
\begin{ruledtabular}
\begin{tabular} {llllllll}
& \multicolumn{5}{c} {Y$_2$Os$_2$O$_7$} & \multicolumn{2}{c} {Ho$_2$Os$_2$O$_7$}\\[0.5ex]
Temperature (K)	& 250 & 100 & 25 & 10 & 3.5 & 60 & 4\\
\cline{2-6} \cline{7-8}\\[0.2ex]
$a$ ({\AA}) & 10.20425(8) & 10.19766(27) & 10.19676(27) & 10.19647(27) & 10.19657(27) & 10.20512(15) & 10.20490(17)\\
$V$ ({\AA}$^3$) & 1062.536(15) & 1060.479(49) & 1060.197(49) & 1060.106(49) & 1060.138(49) & 1062.808(28) & 1062.737(30)\\
$x$(O1) & 0.33510(21) & 0.33541(17) & 0.33546(17) & 0.33547(17) & 0.33550(17) & 0.33565(13) & 0.33575(14)\\
$B \rm_{iso}$(Y) ({\AA}$^2$) & 0.558(62) & 0.420(54) & 0.366(53) & 0.368(53) & 0.370(54) & 0.127(59) & 0.124(63)\\
$B \rm_{iso}$(Os) ({\AA}$^2$) & 0.139(39) & 0.098(32) & 0.101(32) & 0.090(32) & 0.100(32) & 0.106(39) & 0.053(42)\\
$B \rm_{iso}$(O1) ({\AA}$^2$) & 0.708(47) & 0.625(39) & 0.625(38) & 0.614(39) & 0.625(39) & 0.287(36) & 0.308(39)\\
$B \rm_{iso}$(O2) ({\AA}$^2$) & 0.276(116) & 0.220(94) & 0.272(95) & 0.242(95) & 0.236(96) & 0.287(36) & 0.308(39)\\
Os-O1 ({\AA}) & 2.0020(14) & 2.0021(8) & 2.0021(12) & 2.0021(8) & 2.0023(12) & 2.0046(7) & 2.0050(7)\\
O1-Os-O1 (deg) & 81.56(7) & 81.45(6) & 81.44(9) & 81.43(9) & 81.42(9) & 81.37(8) & 81.34(9)\\
 & 98.44(7) & 98.55(6) & 98.56(9) & 98.57(9) & 98.58(9) & 98.63(8) & 98.66(9)\\
Os-O1-Os (deg) & 128.59(17) & 128.42(15) & 128.40(15) & 128.39(15) & 128.38(16) & 128.30(14) & 128.25(15)\\
$R \rm_{Bragg}$ & 4.14\% & 4.15\% & 4.48\% & 4.47\% & 4.73\% &  4.44\% & 5.97\%\\
\end{tabular}
\end{ruledtabular}
\end{table*}

\begin{figure}
\includegraphics[clip,width=8.5cm]{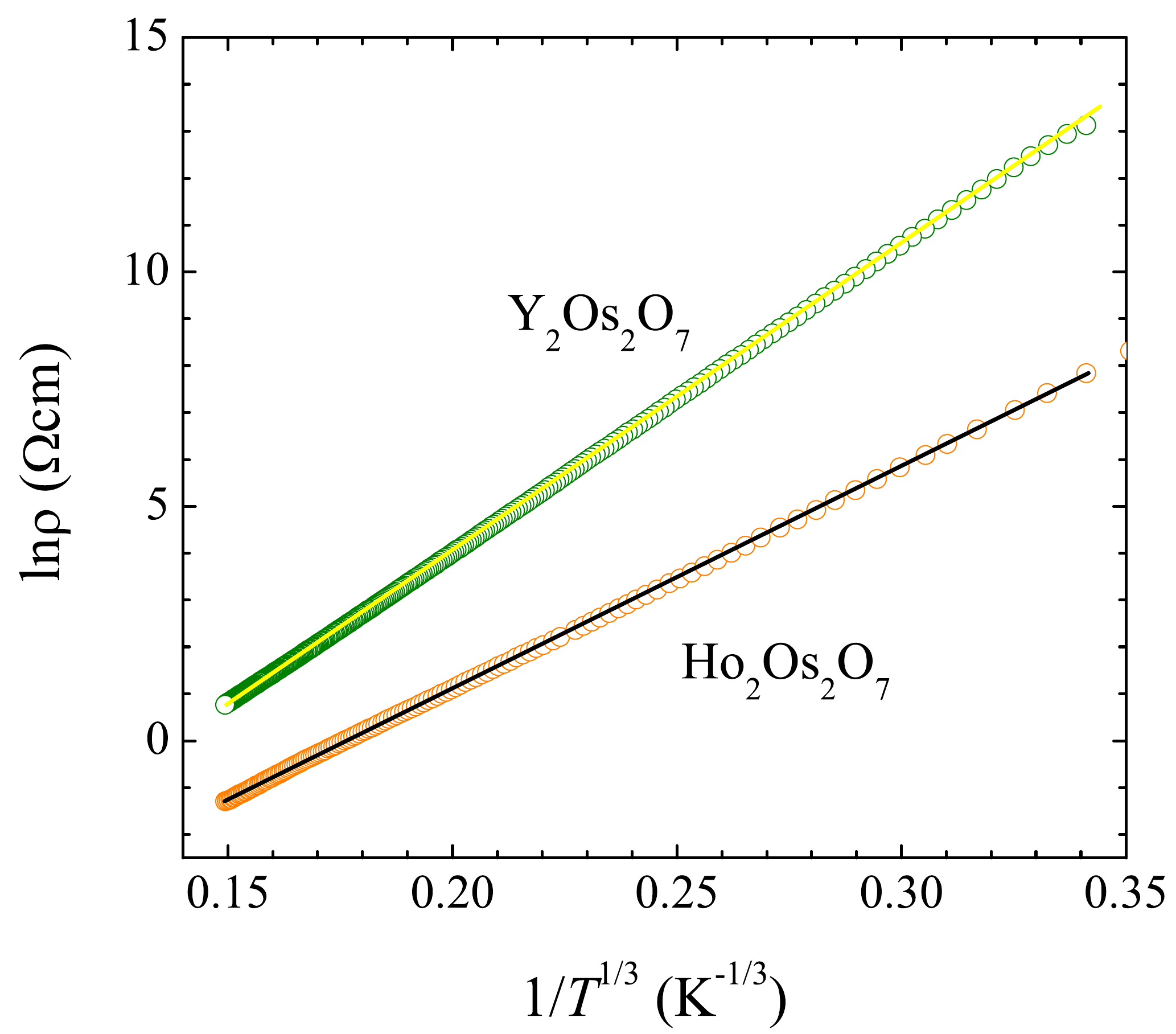}
\caption{(Color online) Temperature dependence of electrical resistivity below room temperature in zero magnetic field. Solid lines are 2D VRH fittings.}
\label{RT}
\end{figure}

The temperature dependence of electrical resistivity was measured on one piece of dense pellet prepared by cold-pressing and further heat treatment. As shown in Fig. \ref{RT},  both compositions are insulators below room temperature. The temperature dependence of electrical resistivity can be well described by the two-dimensional (2D) variable range hopping (VRH) model $\rho = \rho_0$exp($T_0/T^{1/4}$).

\begin{figure}
\includegraphics[clip,width=8.5cm]{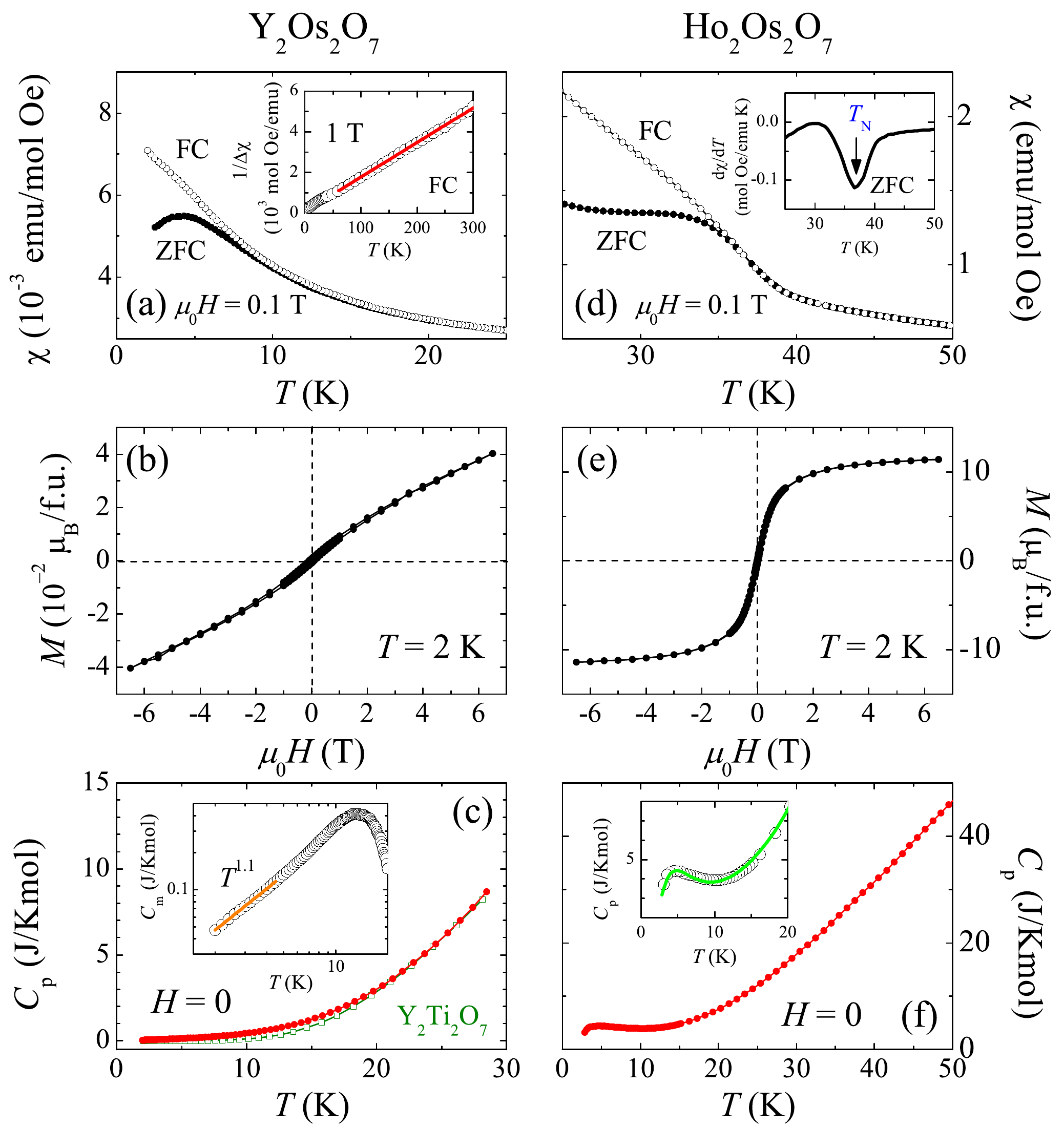}
\caption{(Color online) The DC magnetic susceptibility, magnetization, and specific heat of (a-c) Y$_2$Os$_2$O$_7$ and (d-f) Ho$_2$Os$_2$O$_7$. The red solid line in the inset of (a) is the Curie-Weiss fit and $\Delta \chi = \chi - \chi_0$. $C \rm_p$ of Y$_2$Ti$_2$O$_7$ in (c) is scaled by multiplying a factor of 0.9 to match the high-temperature phonon specific heat of Y$_2$Os$_2$O$_7$. Inset of (c): magnetic specific heat of Y$_2$Os$_2$O$_7$. The solid line is a linear fit. Inset of (d): the derivative of the ZFC susceptibility. Inset of (f): specific heat with a two-level Schottky fit (solid line) at low temperatures.}
\label{DC}
\end{figure}

The magnetic susceptibility  measured with $\mu_0H$ = 0.1 T of Y$_2$Os$_2$O$_7$ (Fig. \ref{DC}(a)) shows a splitting between the field-cooling (FC) and zero-field-cooling (ZFC) curves below 10 K. Furthermore, the ZFC curve shows a broad peak around 4 K. The Curie-Weiss fitting $\chi = \chi_0 + C/(T-\theta_{\rm{CW}})$ ($\chi_0$ is a temperature-independent term) of the inverse susceptibility measured with $\mu_0H$ = 1.0 T between 60 K and 300 K (insert of Fig. \ref{DC}(a)) yields a Curie-Weiss temperature $\theta_{\rm{CW}}$ = -4.8(3) K and an effective moment $\mu \rm_{eff}$ = 0.49(2) $\mu \rm_B$/Os$^{4+}$. This negative $\theta_{\rm{CW}}$ suggests an antiferromagnetic (AFM) interaction among the Os$^{4+}$ spins. The obtained $\mu \rm_{eff}$ is much smaller than the expected 2.83 $\mu_{\rm B}$/Os$^{4+}$ for an $S$ = 1 system, supporting that the SOC indeed strongly suppresses the Os moment in Y$_2$Os$_2$O$_7$. The magnetization measured at 2 K for Y$_2$Os$_2$O$_7$ (Fig. \ref{DC}(b)) shows a weak hysteresis loop.  The specific heat, $C \rm_p$, of Y$_2$Os$_2$O$_7$ shows no anomaly related to the possible long-range magnetic ordering down to 2 K. However, its magnetic specific heat $C \rm_{mag}$ obtained by using $C \rm_p$ of Y$_2$Ti$_2$O$_7$ \cite{Y2Ti2O7} as a phonon reference shows a broad peak at 13 K, following a $T^{1.1}$ dependence below 4.5 K (Fig. \ref{DC}(c)). The calculated magnetic entropy is 0.48 J/Kmol, and is only about 5.3\% of $R$ln3 for a magnetic ordering with $S$ = 1. All the above results, including the ZFC/ FC splitting, the broad peak of the ZFC data, the weak hysteresis loop, and the linear-$T$ dependence of $C \rm_{mag}$, fit to the characteristic behaviors of a SG system at low temperatures. \cite{Mydosh_2} The susceptibility and specific heat demonstrate that the Os$^{4+}$ moments start to develop short-range correlations around 10 K and enter a glassy state below 4 K.

For Ho$_2$Os$_2$O$_7$, the magnetic susceptibility shows a sharp increase below 40 K and a splitting of the ZFC and FC data below 36 K (Fig. \ref{DC}(d)). This sharp increase is absent in Y$_2$Os$_2$O$_7$. The transition temperature $T \rm_N$ = 36 K is defined by the peak position of the derivative of the ZFC data.  Here, the Curie-Weiss fitting cannot deduce a reasonable effective moment of Os$^{4+}$ ions since the high-temperature susceptibility is dominated by the paramagnetic Ho$^{3+}$ ions. \cite{Ho2Ru2O7_2} However, it is reasonable to believe that the effective moment of Os$^{4+}$ in Ho$_2$Os$_2$O$_7$ is similar to that in Y$_2$Os$_2$O$_7$. A weak hysteresis similar to Y$_2$Os$_2$O$_7$ is observed in $M(H)$. As shown from the magnetization (Fig. \ref{DC}(e)), the moment of Ho$_2$Os$_2$O$_7$ at 6.5 T is about half of the saturation moment of Ho$^{3+}$ ($\mu \rm_{sat}$ = 10.60 $\mu \rm_B$/Ho$^{3+}$). This is a characteristic behavior for the Ising spin anisotropy in the Ho-pyrochlores. \cite{Pyrochlore} The specific heat of  Ho$_2$Os$_2$O$_7$ shows no distinct anomaly but a hump around 4 K, which can be described by a two-level Schottky model with a gap about 10.5(1) K (inset of Fig. \ref{DC}(f)). This could be attributed to the splitting of the Ho ground doublets associated with the onset of magnetic correlations, which has been observed in Yb$_2$Sn$_2$O$_7$. \cite{Yb2Sn2O7} The absence of any distinct anomaly in $C \rm_p$($T)$ suggests that the sharp increase and ZFC/FC splitting in the magnetic susceptibility curves result from a magnetic transition of Os$^{4+}$ ions, because (i) the rare-earth spins usually order at temperatures lower than 10 K; (ii) if the Ho$^{3+}$ spins indeed order, the large moment should induce a distinct change in the specific heat. On the other hand, it is difficult to observe the small entropy change related to the small moment of Os$^{4+}$ spins on top of a large phonon background.

\begin{figure}
\includegraphics[clip,width=6.5 cm]{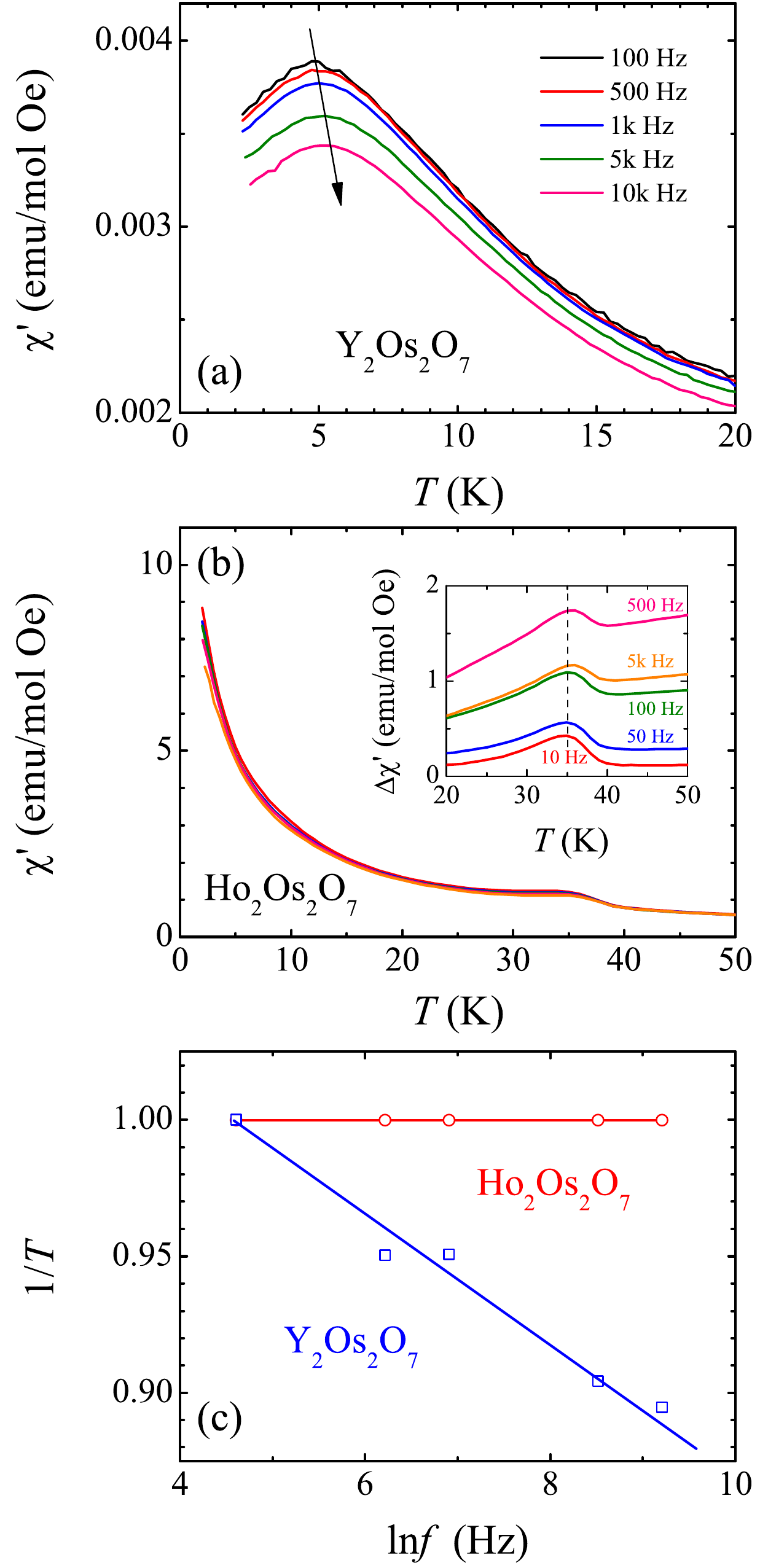}
\caption{(Color online) Real part of the AC susceptibility ($\chi'$) of (a) Y$_2$Os$_2$O$_7$ and (b) Ho$_2$Os$_2$O$_7$ measured with $H \rm_{ac}$ = 12 Oe and $H \rm_{dc}$ = 0. Inset: $\chi'$ of the Os$^{4+}$ moments in Ho$_2$Os$_2$O$_7$. Contribution from the Ho$^{3+}$ moments is subtracted from the total after fitting the Curie-Weiss tail below 6 K. (c) The normalized frequency dependencies of the inverse of the transition temperatures for both compounds.}
\label{acms}
\end{figure}

To further resolve the magnetic ground states of Y$_2$Os$_2$O$_7$ and Ho$_2$Os$_2$O$_7$, we perform AC susceptibility and neutron powder diffraction measurements. The real part of the AC susceptibility ($\chi'$) for Y$_2$Os$_2$O$_7$ shows a frequency-dependent peak around 5 K (Fig. \ref{acms}(a)). This peak shifts to lower temperatures and its intensity increases as the frequency $f$ of the excitation field decreases. This behavior is a typical feature of the dynamics of a spin-glass system. The inverse of the peak temperature $T \rm_{SG}$ is approximately linear with ln$f$ and satisfies the relation $f = f_0\exp(-\Delta/T \rm_{SG})$ with an energy barrier $\Delta = 204(18)$ K (Fig. \ref{acms}(c)). The Mydosh parameter $\Delta T \rm_{SG}$/[$T\rm_{SG} \Delta \rm{log}$$f$], which is a quantitative measure of the frequency shift, is estimated to be 0.022(2). This is near the expected range of 0.004-0.018 for conventional spin-glass systems. \cite{Mydosh_2} For Ho$_2$Os$_2$O$_7$, $\chi'$ shows a transition around 35 K (Fig. \ref{acms}(b)), which is consistent with the transition temperature observed from the DC susceptibility. To study better its nature, a background estimated by using the Curie-Weiss fitting of the data below 6 K is subtracted from the total $\chi'$. Here we assume this background comes from the paramagnetism of the Ho$^{3+}$ ions. The obtained result (inset to Fig. \ref{acms}(b)) clearly shows that the transition temperature is frequency independent, suggesting a long-range magnetic ordering around 36 K for Ho$_2$Os$_2$O$_7$.

\begin{figure}
\includegraphics[clip,width=7.0cm]{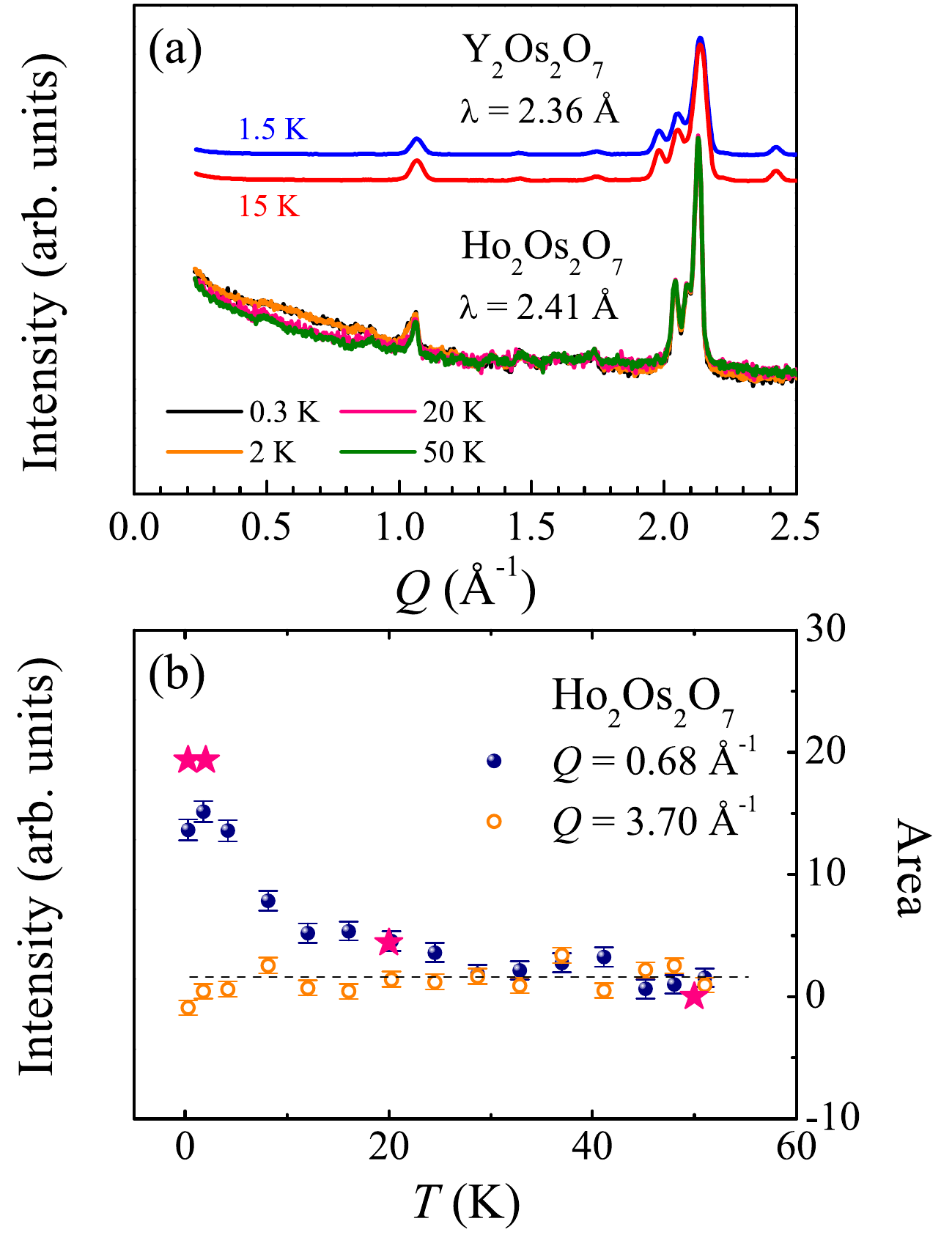}
\caption{(Color online) (a) Neutron diffraction patterns of Y$_2$Os$_2$O$_7$ collected on HB-1A with $\lambda$ = 2.36 {\AA} and Ho$_2$Os$_2$O$_7$ collected on HB-2A with $\lambda$ = 2.41 {\AA}. The intensity of Y$_2$Os$_2$O$_7$ patterns are shifted for clarify. The unit of intensity is in a log scale. (b) Temperature dependence of the intensity for Ho$_2$Os$_2$O$_7$ at $Q$ = 0.68 {\AA}$^{-1}$ ($2\theta = 15^{\circ}$) (solid circles) and $Q$ = 3.7 {\AA}$^{-1}$ ($2\theta = 90^{\circ}$) (open circles). The scaled integration of the diffuse scattering within 0.22 $\leq Q \leq$ 1.3 {\AA}$^{-1}$ is also plot on the right scale as the star symbols. The dashed line is a guide for the eyes, illustrating that the diffuse scattering starts to increase below 30 K.}
\label{neutron}
\end{figure}

Figure \ref{neutron}(a) shows the neutron powder diffraction patterns of Y$_2$Os$_2$O$_7$ collected at 1.5 and 15 K. There is no extra reflection or abnormal peak intensity change observed, which agrees with a spin-glass ground state below 4 K. Diffraction patterns of Ho$_2$Os$_2$O$_7$ collected below 50\,K are also present in Fig. 4(a). No extra reflections or abnormal peak intensity change was observed down to 0.3 K. This seems to suggest the absence of a long-range magnetic ordering of both Ho and Os sublattices, which is against the magnetic measurements. However, a significant temperature-dependent diffuse scattering is observed at low $Q$ as highlighted in Fig. \ref{neutron}(b).  Compared to the $T$-independent background at high $Q$ = 3.7 {\AA}$^{-1}$ ($2\theta$ = 90$^{\circ}$), the intensity at $Q$ = 0.68 {\AA}$^{-1}$ ($2\theta$ = 15$^{\circ}$) gradually increases below 30 K, which is around the transition temperature associated with the Os sublattice ($\sim$ 36 K defined from magnetic measurements). By subtracting the 50 K data as a background, the diffuse scattering at low $Q$ was integrated. The resulting integrated area (labeled as star symbols in Fig. \ref{neutron}(b)), which could be employed to characterize the magnitude of the diffuse scattering, also follows the trend of the diffuse scattering intensity at $Q$ = 0.68 {\AA}$^{-1}$. Similar temperature-dependent diffuse scattering signals at low $Q$ was also observed in other Ho$_2M_2$O$_7$ ($M$ = Ti, Sn, Ru) pyrochlores, \cite{Ho2Ti2O7,Ho2Sn2O7,Ho2Ru2O7} and was attributed to the short-range correlation of Ho$^{3+}$ spins. This diffuse scattering in neutron diffraction pattern usually occurs at much lower temperatures in Ho$_2M_2$O$_7$ with non-magnetic $M$ ions. For example, Ho$_2$Ti$_2$O$_7$ starts to develop a short-range correlation below 2 K. \cite{Ho2Ti2O7} However, in Ho$_2$Ru$_2$O$_7$, where Ru$^{4+}$ spins order below 95 K, the diffuse scattering appears just below the N\'eel temperature due to the Ho-Ru coupling. \cite{Ho2Ru2O7} Therefore, the diffuse scattering appearing at as high as 30 K in Ho$_2$Os$_2$O$_7$ suggests (1) there is no long-range ordering of Ho sublattice down to 0.3 K, (2) Os$^{4+}$ is not in a non-magnetic state as expected by atomic physics in the presence of strong SOC, and (3) Os sublattice orders magnetically below 36 K which induces the diffuse scattering of Ho sublattice via Ho 4$f$-Os 5$d$ interaction. The absence of extra reflections or peak intensity change in our neutron diffraction patterns suggests the ordered moment should be small. This is supported by the temperature dependence of specific heat which is featureless around 36 K. It is noteworthy that the entropy change  across the magnetic ordering in Sr$_2$YIrO$_6$ was reported to be small. \cite{Cao_SrYIrO} Single crystal neutron diffraction study might be able to resolve the ordering pattern of Os sublattice in Ho$_2$Os$_2$O$_7$ .

Obviously, magnetism was observed in both compounds investigated in this work. The SG state with weak magnetism in Y$_2$Os$_2$O$_7$ signals the importance of SOC. This could be better illustrated by comparing Y$_2$Os$_2$O$_7$ with an isostructural material Y$_2$Ru$_2$O$_7$ where the magnetic Ru$^{4+}$ has a $4d^4$ electron configuration. At 100 K, the lattice parameters are $a$ = 10.1977(3) {\AA} and $a$ = 10.116(3) {\AA} for Y$_2$Os$_2$O$_7$ and Y$_2$Ru$_2$O$_7$, \cite{Y2Ru2O7_refinement} respectively. The O(1) position $x$, which determines the trigonal distortion of octahedra, is 0.3354(2) for Y$_2$Os$_2$O$_7$ and 0.335(1) for Y$_2$Ru$_2$O$_7$, \cite{Y2Ru2O7_refinement} respectively. Despite the structural similarity, their magnetic properties are drastically different. Y$_2$Ru$_2$O$_7$ orders antiferromagnetically below $T_{\text N} = 76$ K with a very large $\theta_{\text{CW}} = -1250$ K \cite{Y2Ru2O7_Susceptibility,Y2Ru2O7_Cp} and the magnetic moment $\mu_{\text{eff}}$ = 3.1 $\mu_{\text B}$/Ru$^{4+}$ is close to the expectation for an $S=1$ system. Apparently, the SE dominates over SOC in Y$_2$Ru$_2$O$_7$ and drives magnetic ordering. For Os$^{4+}$ ions in Y$_2$Os$_2$O$_7$, SOC is certainly much stronger and a singlet ground state is expected. The fact that Y$_2$Os$_2$O$_7$ has a small $\theta_{\text{CW}} = - 4.8$ K and a $\mu_{\text{eff}}$ = 0.49 $\mu_{\text B}$/Os$^{4+}$  signals the importance of the intersite superexchange interactions and trigonal crystal fields. Although the superexchange may still favor magnetic ordering, the frustration of the underlying lattice and/or the disorder could then drive possible ordering into a glassy state. The weak magnetism in Y$_2$Os$_2$O$_7$ suggests that Y$_2$Os$_2$O$_7$ stays close to the quantum phase transition between the non-magnetic state in the strong SOC limit and the magnetic state in the strong superexchange limit.

\begin{figure}
\includegraphics[clip,width=8cm]{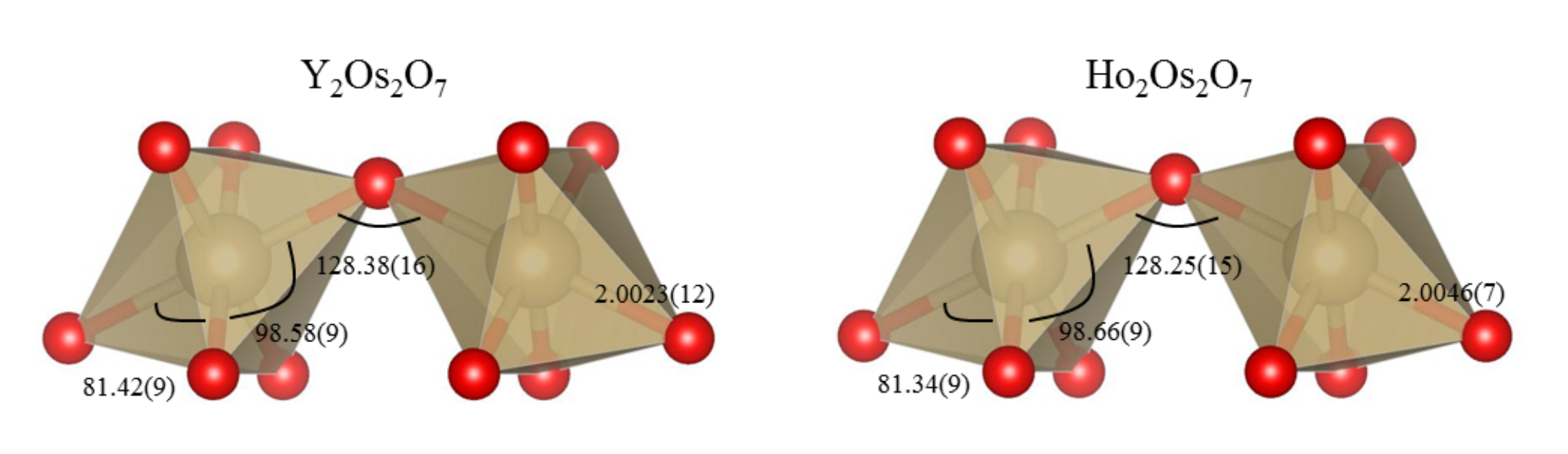}
\caption{(Color online) Comparison of local distortion between Y$_2$Os$_2$O$_7$ and Ho$_2$Os$_2$O$_7$ around 4 K. OsO$_6/2$ octahedra are connected by sharing the oxygen (red) and Os (yellow) lies in the center of the octahedron.}
\label{local}
\end{figure}

As the consequence of proximity to the phase transition, the SG ground state of Y$_2$Os$_2$O$_7$ is fragile and should be easily affected by perturbations. The comparison between Y$_2$Os$_2$O$_7$ and Ho$_2$Os$_2$O$_7$ highlights this point. With the similar ionic radius for Ho$^{3+}$ (IR = 1.015 {\AA} ) and Y$^{3+}$ (IR= 1.019 {\AA}), \cite{IR}  Y$_2$Os$_2$O$_7$ and Ho$_2$Os$_2$O$_7$ show nearly identical structural features, as shown in Fig. \ref{local}. Around 4 K, the lattice parameter is $a$ = 10.2049(2) {\AA} and the O(1) position is 0.3358(1) for Ho$_2$Os$_2$O$_7$, close to $a$ = 10.1966(3) {\AA} and 0.3355(2) for Y$_2$Os$_2$O$_7$. The detailed structural information for both compounds at different temperatures are summarized in Table 1. The structural similarity motivates us to consider the importance of $f-d$ interaction in stabilizing the long-range magnetic ordering of Os sublattice in Ho$_2$Os$_2$O$_7$. A previous work on pyrochlore iridates has shown that $f$-$d$ interaction is likely to stabilize the magnetic ordered state. \cite{Chen12} Since the 4$f$ electrons are highly localized, the $4f-5d$ interaction is expected to be weak. This, in turn, signals that the singlet ground state magnetism of Os sublattice in $R_2$Os$_2$O$_7$ is rather fragile. It will be interesting to tune the magnetic ground states by applying external stimuli such as pressure and/or magnetic fields.

\section{Conclusions}

In summary, we have studied the magnetic properties of two isostructural pyrochlore osmates. Y$_2$Os$_2$O$_7$ shows a spin glass state. However, Os sublattice in Ho$_2$Os$_2$O$_7$ orders magnetically with $T \rm_N$ = 36\,K. The sharp difference highlights that the singlet ground state magnetism in pyrochlore osmates is fragile and can be tuned by the weak $4f-5d$ interaction. Similar to that in Sr$_2$YIrO$_6$, the entropy change associated with the magnetic ordering is small. Although more materials should be studied before coming to a conclusion, a fragile magnetic ground state and a small entropy change across T$_N$ seem to be two general features for the singlet ground state magnetism in transition metal oxides with strong SOC.

Considering the magnetic ordering of Os sublattice in Ho$_2$Os$_2$O$_7$ is induced by the weak $4f-5d$ interaction, the magnitude of rare earth moment is expected to play an essential role in determining the magnetic ordering temperature of Os sublattice in \textit{R}$_2$Os$_2$O$_7$ (\textit{R} = rare earth). A detailed study of the magnetism in the whole \textit{R}$_2$Os$_2$O$_7$ system is in progress and our preliminary results support the above statement. Meanwhile, the fragile magnetic ground state is expected to be sensitive to external stimuli such as pressure and/or magnetic fields. Studies under high pressure or in high magnetic fields are desired.

\section{Acknowledgements}

Work at ORNL was supported by the US Department of Energy, Office of Science, Basic Energy Sciences, Materials Sciences and Engineering Division. Z.Y.Z. and N.T. acknowledge the CEM, and NSF MRSEC, under grant DMR-1420451. Research conducted at ORNL's HFIR was sponsored by the Scientific User Facilities Division, Office of Basic Energy Sciences, and US Department of Energy. H.D.Z. thanks the support from NSF-DMR-1350002. D.G.M. acknowledges support from the Gordon and Betty Moore Foundations EPiQS Initiative through Grant GBMF4416.

\end{document}